\documentclass[onecolumn]{aastex63}
\usepackage{amsmath}
\usepackage{multirow}
\usepackage{hyperref}
\newcommand{\PMO}{Key Laboratory of Dark Matter and Space Astronomy, Purple Mountain Observatory, Chinese Academy of Sciences, Nanjing 210033, People's Republic of China}
\newcommand{\USTC}{School of Astronomy and Space Science, University of Science and Technology of China, Hefei, Anhui 230026, People's Republic of China}
\newcommand{\NNU}{Department of Physics and Institute of Theoretical Physics, Nanjing Normal University, Nanjing 210046, People's Republic of China}
\newcommand{\ITGU}{Institut f\"{u}r Theoretische Physik, Goethe Universit\"{a}t, Max-von-Laue-Str. 1, D-60438 Frankfurt am Main, Germany.}

\begin{document}
\title{The Bulk Properties of Isolated Neutron Stars Inferred from the Gravitational Redshift Measurements}
\author[0000-0003-3471-4442]{Chuan-Ning Luo}
\author[0000-0001-9120-7733]{Shao-Peng Tang}
\affil{\PMO}
\affil{\USTC}
\author[0000-0002-9078-7825]{Jin-Liang Jiang}
\affil{\PMO}
\affil{\ITGU}
\author[0000-0001-7821-3864]{Wei-Hong Gao}
\email{Corresponding author.~gaoweihong@njnu.edu.cn}
\affil{\NNU}
\author[0000-0002-9758-5476]{Da-Ming Wei}
\email{Corresponding author.~dmwei@pmo.ac.cn}
\affil{\PMO}
\affil{\USTC}

\begin{abstract}
    The measurements of the bulk properties of most isolated neutron stars (INSs) are challenging tasks. Tang et al. (2020) have developed a new method, based on the equation of state (EoS) of neutron star (NS) material constrained by the observational data, to infer the gravitational masses of a few INSs whose gravitational redshifts are available. However, in that work, the authors only considered the constraints on the EoS from nuclear experiments/theories and the gravitational wave data of GW170817; the possible phase transition has not been taken into account. In this work, we adopt three EoS models (including the one incorporates a first-order strong phase transition) that are constrained by the latest multimessenger NS data, including in particular the recent mass\textendash radius measurements of two NSs by Neutron Star Interior Composition Explorer, to update the estimation of the gravitational masses of RBS 1223, RX J0720.4-3125, and RX J1856.5-3754. In comparison to our previous approach, the new constraints are tighter, and the gravitational masses are larger by about $0.1M_\odot$. All the inferred gravitational masses are within the range of the NS masses measured in other ways. We have also calculated the radius, tidal-deformability, and moment of inertia of these sources. The inclusion of the first-order strong phase transition has little influence on modifying the results.
\end{abstract}

\section{Introduction}
The mass determination of compact stars, the leftover products of stellar death, is fundamental in the researches of astronomy. As one of the basic quantities one can measure, the mass of neutron star (NS) may enable us to reveal the formation and evolution mechanism of stars, since different progenitors and evolution channels can leave different imprints on their final masses. For example, \citet{2013ApJ...778...66K} found that the mass distribution of NSs in NS\textendash white dwarf (NSWD) systems is peaked at heavier mass than that of NSs in binary NS systems, indicating significant mass accretion of NSs in NSWD via the so-called recycling process. Meanwhile, mass distribution of NSs can also be used to probe the currently unknown maximum mass of NS, which is important for determining the equation of state (EoS) of ultra-dense matter and whether stars collapse into NSs or black holes \citep{Alsing2018, Shao2020}. On the other hand, the birth mass of NS is also fascinating and can be used to check the theoretical expectations for remnants mass-produced by electron-capture versus Fe-core collapse SNe \citep{2004ApJ...612.1044P, 2013ApJ...778...66K}. 

Isolated NSs (INSs) are believed to trace initial masses when they were born. However, the mass measurement for INSs is much more challenging than NSs in binary systems \citep{2016ARA&A..54..401O}. There are some promising methods for such mass measurement: (1) the pulse profile modeling of the emission from the hotspots of NS, but the targets of the Neutron Star Interior Composition Explorer (NICER) mission currently have masses that are likely larger than their initial birth masses; (2) the glitches (sudden and temporary change in the NS spin) of some young (isolated) pulsars can be used to put constraints on the mass of NS \citep{2017NatAs...1E.134P}, but such a method may provide rather loose constraints; (3) the mass of INS with gravitational redshift measurement can be extracted by taking the advantage of EoS constraining results \citep{Tang2020}, which is benefited from the fact that most of the EoSs tend to give a unique map between the mass and the gravitational redshift of the NS. 

In our previous work of \citet{Tang2020}, the possibility of phase transition (PT) was not considered, and the constraints on EoSs mainly come from the nuclear constraints and gravitational wave (GW) data of GW170817. Since then, some important progresses have been made. Firstly, mass\textendash radius measurements of two NSs have been successfully carried out by the NICER mission, i.e., PSR J0030+0451 \citep{Miller2019, Riley2019} in the low-mass region and PSR J0740-6620 \citep{Miller2021, Riley2021} in the high-mass region. This means more stringent constraint can be made on the EoS, and thus the masses of the other INSs that only have redshift measurement. Secondly, a new approach embedding both the PT and no phase transition (NPT) model as one has been proposed by \citet{Tang2021}, whose method can be used to take into analysis when estimating the mass from the gravitational redshift. 

In view of these new progresses, we carry out this work, focusing on the mass estimation method proposed by \citet{Tang2020}, and update the previous analysis by taking into account the possibility of PT in our EoS models and incorporating the constraints on EoS from the NICER's observations (i.e., the mass\textendash radius measurement of PSR J0030+0451 and PSR J0740-6620) additional to the nuclear constraints and GW data of GW170817. In this work, we estimate the mass, radius, tidal-deformability, and the moment of inertia of the sources RBS 1223, RX J0720.4-3125, and RX J1856.5-3754 with different models. Each of these sources has its gravitational redshift been measured.

Our work is organized as follows. In Section \ref{sec:methods} we introduce the methods. The results of the calculation are presented in Section \ref{sec:results}. Section \ref{sec:discussion} is the conclusion and discussion.

\section{Methods}
\label{sec:methods}
\subsection{The neutron star EoSs constrained with latest multimessenger data}\label{sec:EoS}
Three EoS models, namely the four-pressure (4P) model, the phase transition (PT) model, and the NPT model, have been used in this work. The first so-called four-pressure model \citep{Jiang2020,Tang2020} adopts four pressures $\{P_1, P_2, P_3, P_4\}$ at the corresponding rest-mass densities of $\{1, 1.85, 3.7, 7.4\}\rho_{\rm sat}$ to parameterize the EoS; the second/third are the hybrid parameterization method proposed in \citet{Tang2021}, which is capable to describe generic phenomenological EoS models both with and without a PT. This method constructs the adiabatic indices based on four widely used parameterization models, and the specific expression reads
\begin{equation}\label{eq:Gamma}
    \Gamma(\varepsilon,p,h) = \begin{cases}
                        \Gamma_{\rm crust} & \quad \rho<\rho_0, \\
                        \Gamma_{\rm nuc}(\rho, x) & \quad \rho_0<\rho \leq \rho_1, \\
                        \frac{1}{1+\Upsilon(h,v_k)}\frac{\varepsilon+p}{p} & \quad \rho_1<\rho \leq \rho_2, \\
                        \Gamma_{\rm m} & \quad \rho_2<\rho \leq \rho_2\!+\!\Delta \rho, \\
                        c_{\rm q}^2\frac{\varepsilon+p}{p} & \quad \rho>\rho_2\!+\!\Delta \rho,
                    \end{cases}
\end{equation}
where $\varepsilon$, $p$, $h$, and $\rho$ denote the energy density, the total pressure, the pseudo enthalpy, and the rest-mass density, respectively. $\Gamma_{\rm crust}$ is determined by the tabulated low-density EoS. $\Gamma_{\rm nuc}(\rho, x)$ is calculated under the parabolic expansion-based nuclear empirical parameterization \citep{Steiner2010, Biswas2021}. $\Upsilon(h,v_k)$ is the expansion function used in the causal spectral representation \citep{Lindblom2018}. $\Gamma_{\rm m}$ is the adiabatic index for the piece modeled by a single polytrope \citep{Ozel2009, Read2009}. $c_{\rm q}$ is the sound velocity that describes the high-density part of EoS with constant-speed-of-sound (CSS) parameterization \citep{Alford2013}. For the dividing densities, $\rho_0$ corresponds to the crust-core transition density, and $\rho_1$ is fixed to $1.85\,\rho_{\rm sat}$. The PT and NPT models mainly depend on the parameter $\Gamma_{\rm m}$, and we treat $\Gamma_{\rm m}<1.4$ as the PT model (NPT model otherwise). For PT model, $\rho_2$ means the PT density, and $\Delta \rho$ measures the density jump. While for NPT model, $\rho_2$ and $\rho_2\!+\!\Delta \rho$ are just simply dividing densities.

With the EoS models, we can map EoS parameters to a series of mass\textendash radius (or mass\textendash tidal-deformability) relations \citep{Lindblom2014}; on the contrary, with the observation data of NSs, we can constrain the EoS by Bayesian analysis. The observation data used in the Bayesian inference include the following: the tidal-deformability measurements from GW170817 \citep{2017PhRvL.119p1101A, 2018PhRvL.121p1101A}, the mass\textendash radius measurements of PSR J0030+0451 \citep{Miller2019, Riley2019} and PSR J0740-6620 \citep{Miller2021, Riley2021}. As shown by previous research, results got by using data of \citet{Riley2019}/\citet{Riley2021}, and \citet{Miller2019}/\citet{Miller2021} show good consistency \citep{Raaijmakers2019, Jiang2020}. Thus we only use the data of \citet{Riley2019} for PSR J0030+0451 \footnote{The data of ST+PST case is considered, see \url{http://doi.org/10.5281/zenodo.3386449}} and \citet{Riley2021} for PSR J0740-6620. \footnote{The data file ``STU/NICERxXMM/FI\_H/run10" from \url{https://zenodo.org/record/4697625\#.YKMcuy0tZQJ} is taken into analysis.} For GW data, we use the interpolated, marginalized likelihood of \citet{Vivanco2020}, which shows good consistency with the original GW data analysis. While for data of NICER, we use the Gaussian kernel density estimation (KDE) of the publicly distributed posterior samples of mass (GW) and radius to build the likelihood. The final likelihood is a production of these two parts and is sampled using the PyMultinest \citep{Buchner2016} package.

\begin{figure}[htb]
    \centering
    \gridline{\fig{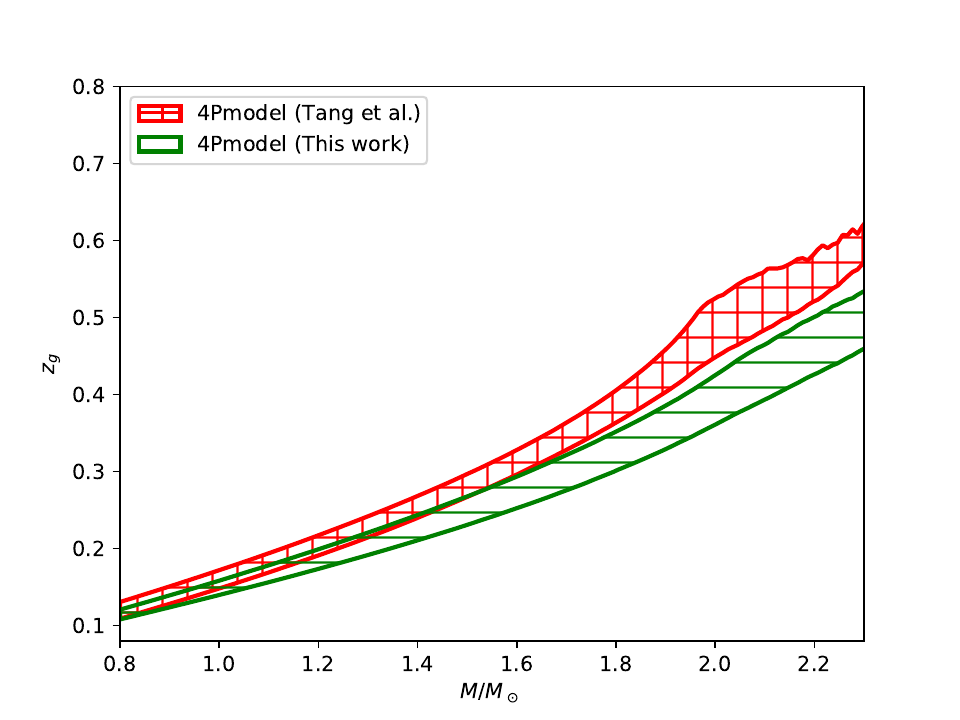}{0.48\textwidth}{(a)}
              \fig{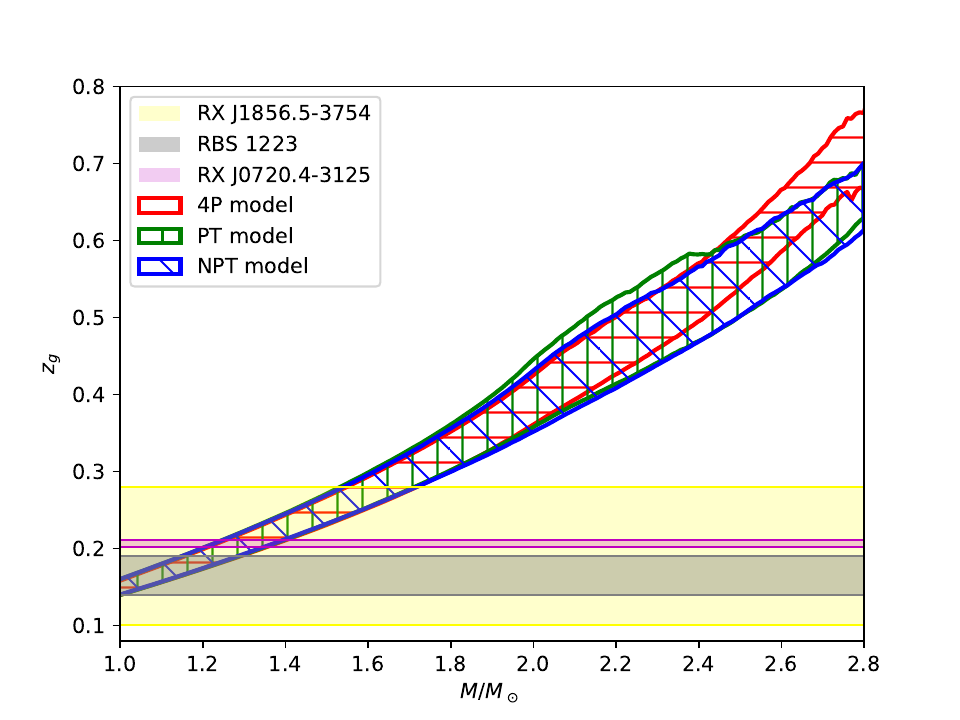}{0.48\textwidth}{(b)}}
    \caption{ Panel (a) and (b) both represent the relationship between the gravitational redshift and the mass of NS. Panel (a) shows the comparison between the results of 4P model in this work (green lines) and that obtained in \citet{Tang2020} (red lines); where $M_{\rm TOV}\leq 2.3M_\odot$ was assumed). The shadow areas represent 68\% confidence region. Panel (b) shows the comparison of results obtained from different models in this work. Red-, green-, and blue-band plots represent the $68\%$ confidence regions of the results obtained from 4P, PT, and NPT model, respectively. Grey-, purple-, and yellow-band plots show the 95\% highest-posterior-density interval of gravitational redshift measurements for sources RBS 1223, RX J0720.4-3125, and RX J1856.5-3754, respectively.}
    \label{fig:z_distribution}
\end{figure}

\subsection{Inferring bulk properties from gravitational redshift measurements}
The gravitational redshifts of three members of the so-called "The Magnificent Seven", RBS 1223, RX J0720.4-3125, and RX J1856.5-3754, have been determined to be $0.16^{+0.03}_{-0.02}$, $0.205^{+0.006}_{-0.003}$, and $0.22^{+0.06}_{-0.12}$ (the 95\% highest-posterior-density interval), respectively, by using the multiepoch observations conduced by XMM-Newton and their X-ray spin phase-resolved spectroscopic studies \citep{Hambaryan2014, Hambaryan2017}.

Given an EoS, we can map the gravitational redshift $z_{\rm g}$ with the compactness $\mathcal{C}$ through relation $z_{\rm g}=1/\sqrt{1-2\mathcal{C}}-1$. Here compactness is defined as $\mathcal{C}=GM/Rc^2$. $c$, $G$, $M$ and $R$ are the speed of light, Newton's gravitational constant, the gravitational mass, and circumferential radius, respectively. Since each posterior EoS sample in our model gives a monotonous relation between the mass and compactness, we can also uniquely find mass and radius by redshift. The tidal-deformability can also be determined if the mass is known\citep{Jiang2019}. Further, the moment of inertia $I$ can be evaluated using an EoS-insensitive relation between tidal-deformability $\Lambda$ and the dimensionless moment of inertia $\bar{I}=c^4I/G^2M^3$. The relation is called I-Love relation \citep{Yagi2013, Landry2018} and reads as follows:
\begin{equation}
    \log_{10}\bar{I}=\sum_{n=0}^{4}a_n(\log_{10}\Lambda)^n,
\label{eq:funcI}
\end{equation}
where $a_{\rm n}$ are fitting coefficients and are taken from \citet{Landry2018}.

\begin{figure}[htb]
    \gridline{\fig{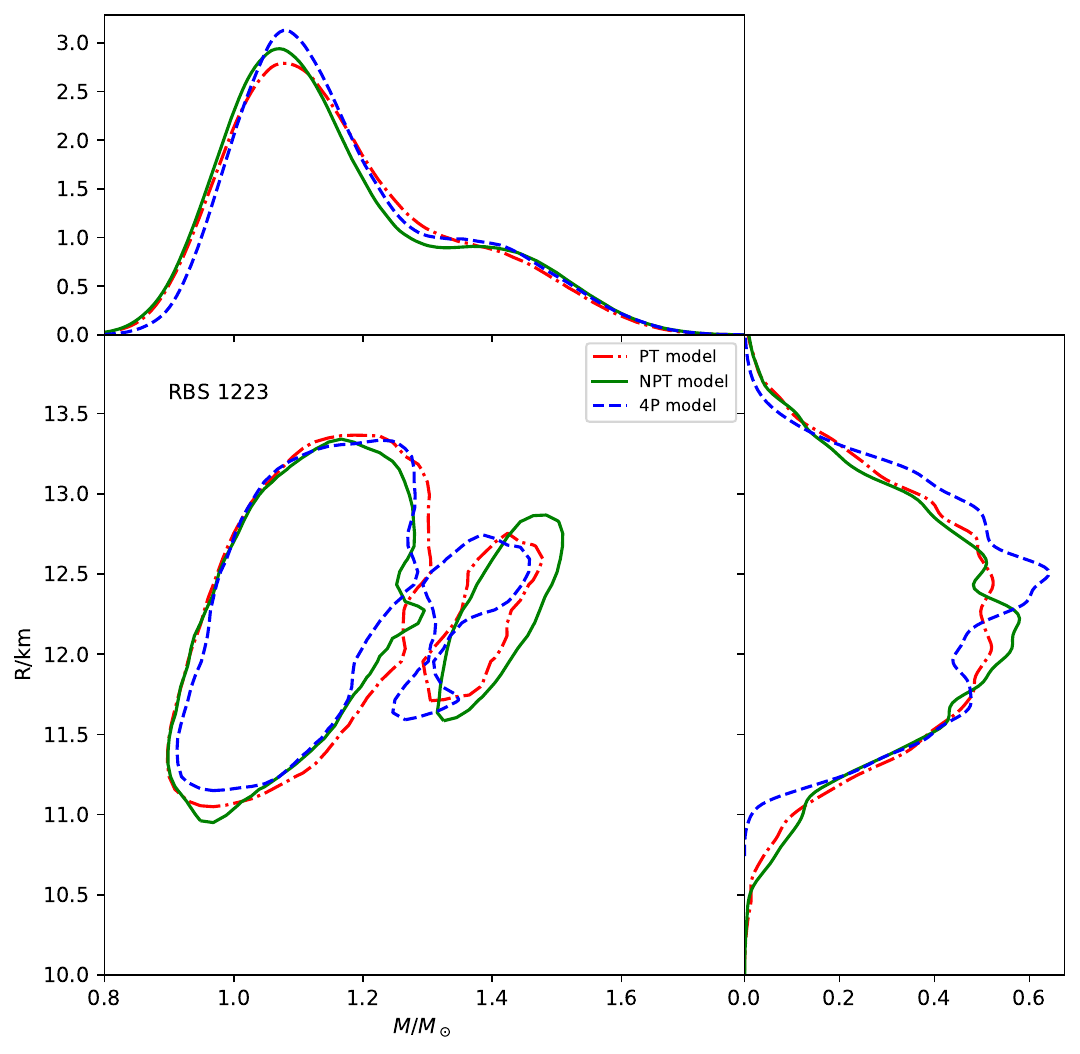}{0.42\textwidth}{(a)}
              \fig{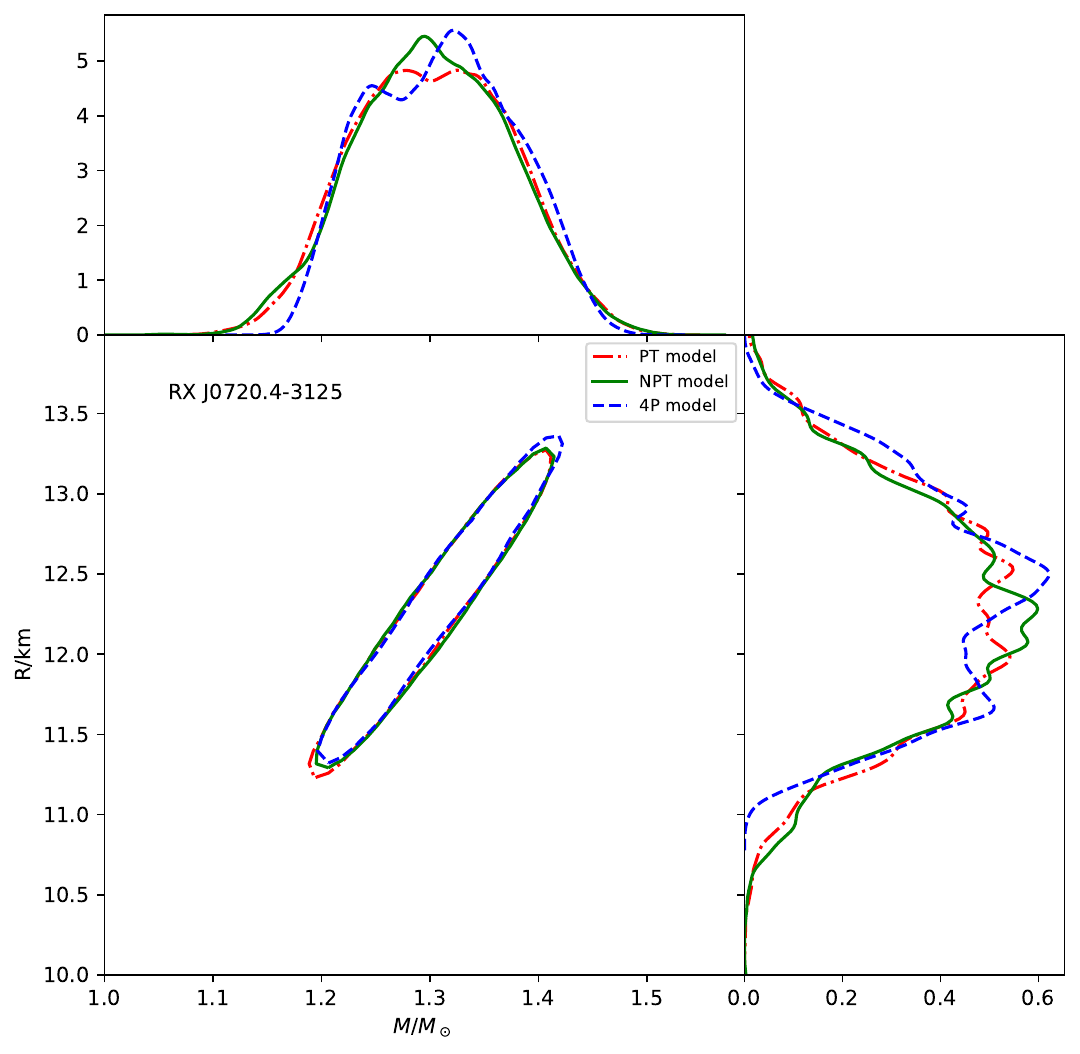}{0.42\textwidth}{(b)}}
    \gridline{\fig{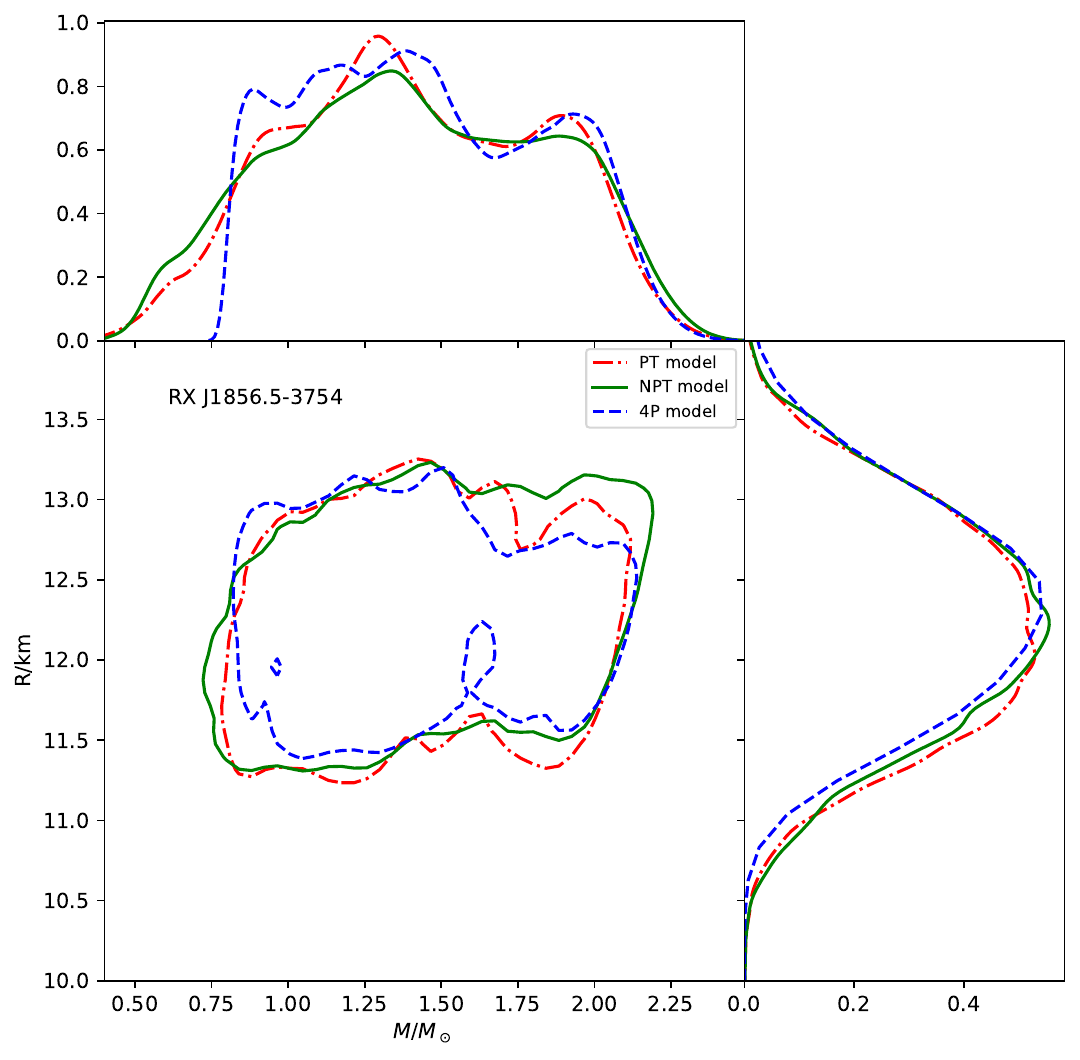}{0.42\textwidth}{(c)}}
    \caption{Interpolated mass\textendash radius distribution for three sources. Panel (a), (b), and (c) represent the results of RBS 1223, RX J0720.4-3125, and RX J1856.5-3754, respectively. Red, green, and blue lines denote the PT, NPT, and 4P model, respectively. In each panel, the contour plots show the 68\% confidence region of the results obtained from different models, while the upper and right sub-graphs show the marginalized probability density function (PDF) of the mass and radius, respectively. }
    \label{fig:ZMR}
\end{figure}

\begin{table}[]
\renewcommand\tabcolsep{8.0pt}
\caption{Results of the mass, radius, tidal deformability, and moment of inertia for the three sources obtained with three EoS models (all in 68\% confidence interval).}
\label{table:result}
\begin{tabular}{|cc|cccc|}
\hline
\multicolumn{2}{|c|}{Source/Parameters}                            & $M/M_{\rm \odot}$      & $R/\rm km$              & $\Lambda$             & $I/10^{45}\rm g\cdot cm^2$             \\ \hline
\multicolumn{1}{|c|}{\multirow{3}{*}{RX J1856.5-3754}} & 4P model  & $1.41_{-0.40}^{+0.50}$ & $12.37_{-0.66}^{+0.68}$ & $420_{-370}^{+3260}$  & $1.60_{-0.62}^{+0.83}$ \\
\multicolumn{1}{|c|}{}                                 & PT model  & $1.38_{-0.42}^{+0.50}$ & $12.21_{-0.68}^{+0.71}$ & $330_{-290}^{+1880}$  & $1.46_{-0.66}^{+0.80}$  \\
\multicolumn{1}{|c|}{}                                 & NPT model & $1.39_{-0.46}^{+0.51}$ & $12.25_{-0.70}^{+0.69}$ & $340_{-290}^{+1960}$  & $1.33_{-0.59}^{+0.96}$  \\ \hline
\multicolumn{1}{|c|}{\multirow{3}{*}{RX J0720.4-3125}} & 4P model  & $1.31_{-0.07}^{+0.07}$ & $12.36_{-0.70}^{+0.65}$ & $641_{-48}^{+56}$     & $1.37_{-0.23}^{+0.26}$  \\
\multicolumn{1}{|c|}{}                                 & PT model  & $1.30_{-0.07}^{+0.08}$ & $12.27_{-0.69}^{+0.71}$ & $507_{-78}^{+77}$     & $1.26_{-0.21}^{+0.24}$  \\
\multicolumn{1}{|c|}{}                                 & NPT model & $1.30_{-0.07}^{+0.07}$ & $12.26_{-0.67}^{+0.68}$ & $509_{-76}^{+81}$     & $1.26_{-0.20}^{+0.24}$  \\ \hline
\multicolumn{1}{|c|}{\multirow{3}{*}{RBS 1223}}        & 4P model  & $1.14_{-0.12}^{+0.23}$ & $12.34_{-0.70}^{+0.60}$ & $1460_{-980}^{+890}$  & $1.13_{-0.24}^{+0.34}$  \\
\multicolumn{1}{|c|}{}                                 & PT model  & $1.13_{-0.13}^{+0.22}$ & $12.23_{-0.70}^{+0.70}$ & $1140_{-740}^{+730}$  & $1.04_{-0.24}^{+0.35}$  \\
\multicolumn{1}{|c|}{}                                 & NPT model & $1.12_{-0.12}^{+0.25}$ & $12.20_{-0.67}^{+0.70}$ & $1120_{-710}^{+750}$  & $1.02_{-0.23}^{+0.35}$  \\ \hline
\end{tabular}
\end{table}

\section{Results}
\label{sec:results}
We compare the results of 4P model in this work and those obtained in \citet{Tang2020}. In the left panel of Fig.~\ref{fig:z_distribution}, we notice that the slope of the new result in mass\textendash redshift relationship is smaller than that of previous work. The main reason is that we have included the measurements from NICER that favor stiffer EoS than the sole GW data. Another reason is that we do not put constraint on the upper limit of $M_{\rm TOV}$ as \citet{Tang2020}, which consequently reduces the slope in the high-mass region. As shown in right panel of Fig.~\ref{fig:z_distribution}, the results of mass\textendash redshift relation constrained with the three EoS models are consistent with each other, especially at the low-redshift region because the EoS has already been constrained well for lower densities. These relations also well cover the ranges of redshift measurement of these sources. Therefore, it is straightforward to simultaneously estimate the mass and radius given the sample of redshift that is randomly drawn from the measured redshift distribution. With the Monte Carlo technique, we can get the corresponding mass and radius distributions for each source, and the results are shown in Table~\ref{table:result} and Fig.~\ref{fig:ZMR}. We notice that the masses and radii of RX J0720.4-3125 are well constrained, benefiting from the narrow distributions of their gravitational redshift measurement. While for RBS 1223, there are two obvious peak structures in its redshift distribution, and the uncertainty is relatively larger. Consequently, its mass distribution inferred from the redshift is broader than RX J0720.4-3125, and exhibits a profile with two lumps. As for RX J1856.5-3754, the uncertainty of its mass becomes even larger, due to its worse constraint on gravitational redshift. Though the inferred masses of the three sources differ from each other, the results of the radius constraints are very similar for all sources, with medium values around $12.3 \rm km$ and uncertainties around $\pm0.7 \rm km$. This is because most of posterior EoS samples constrained by the multimessenger data predict very similar radii for NS in the mass range $1-1.8\,M_{\odot}$.

\begin{figure}[htb]
    \centering
    \includegraphics[width=6in]{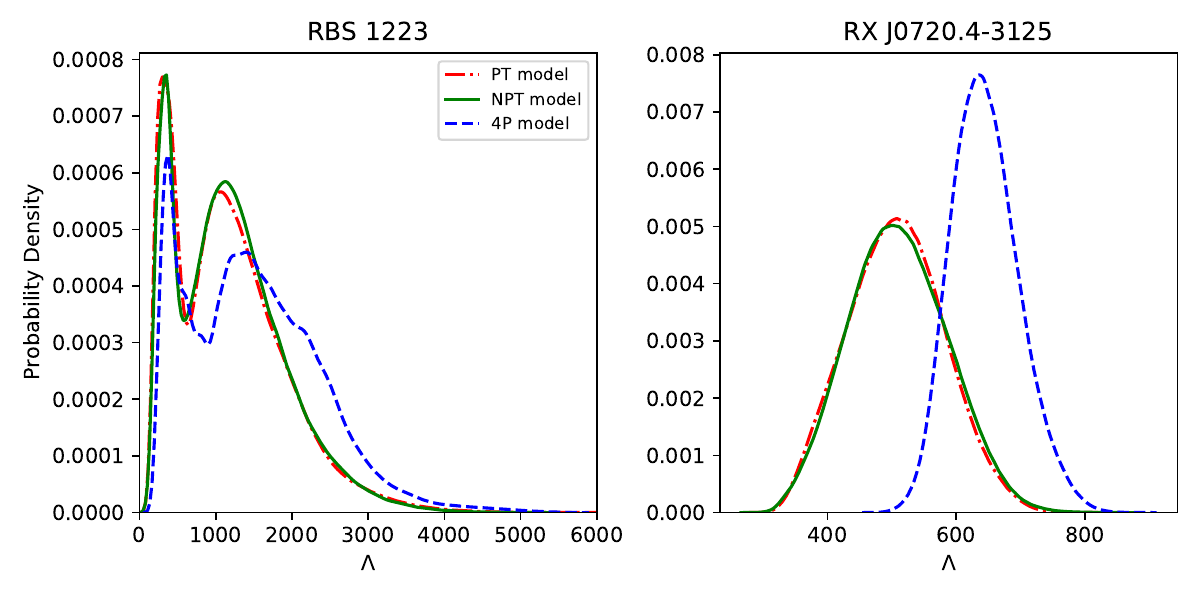}
    \caption{Distribution of tidal-deformability. Red dashed-dotted line, green solid line, and blue dashed line represent the PT, NPT and 4P model, respectively. Left, middle, and right panels show the tidal-deformability of RBS 1223 and RX J0720.4-3125, respectively.}
    \label{fig:tidal_dis}
\end{figure}

\begin{figure}[htb]
    \centering
    \includegraphics[width=6in]{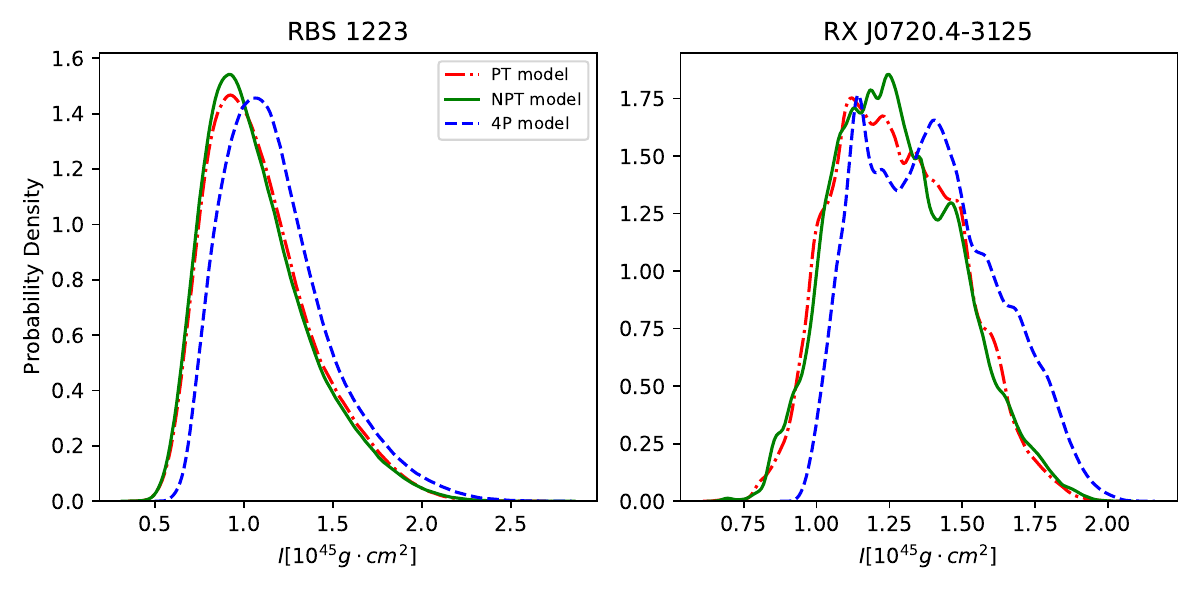}
    \caption{Same as Fig.\ref{fig:tidal_dis} but for the moment of inertia calculated using Eq.~(\ref{eq:funcI}).}
    \label{fig:inertia_dis}
\end{figure}

The distributions of the tidal-deformability ($\Lambda$) are obtained in a similar way as the mass and radius, which are shown in Fig.~\ref{fig:tidal_dis}. We find that the gravitational redshift of the source RX J1856.5-3754 is too inaccurate to give meaningful constraints on $\Lambda$, so we only show results of the other sources. We notice that the tidal-deformability inferred by the 4P model is slightly different from those obtained with the PT and NPT models. In particular, for RX J0720.4-3125, the 4P model predicts a narrower distribution tending to larger values in comparison to the results obtained with the other two models. The reason for such phenomenon is that the $m$\textendash $\Lambda$ relation constrained by the 4P model is more stringent. The distributions of the dimensionless moment of inertia present a similar trend as those of the tidal-deformability since they are directly translated from $\Lambda$ by the universal relation. However, for the moment of inertia $I$ (as shown in Fig.~\ref{fig:inertia_dis}), the difference between 4P model and the other models is smeared out because the distribution of $I$ is dominated by $M$.

We summarize the results of the inferred mass, radius, tidal-deformability, and moment of inertia for each source in Table~\ref{table:result}. Comparing to the previous work \citep{Tang2020}, where the masses of RX J1856.5-3754, RX J0720.4-3125, and RBS 1223 were estimated to be $1.24_{-0.29}^{+0.29}M_{\odot}$, $1.23_{-0.05}^{+0.10}M_{\odot}$, and $1.08_{-0.11}^{+0.20}M_{\odot}$, respectively. The resulting masses inferred in this work become heavier with smaller uncertainties. Because we have incorporated the additional measurement result of PSR J0740-6620, which implicates larger radius than our previous results for the same gravitational mass. This is understandable as larger radius means larger mass for getting the same redshift measured. We also find that there is little difference in all the results between PT model and NPT model. The underlying reason is that INS is not expected to have large mass; therefore the PT onset density. Meanwhile, \citet{2020ApJ...894L...8C} found that strong phase transition below $1.7\rho_{\rm sat}$ ($1\sigma$ level) was ruled out (see also \citealt{2021PhRvD.103f3026T}). Thus we conclude that whether PT is added or not does not affect the results too much. Besides, it should be noticed that the results listed in Table~\ref{table:result} cannot be used to further constrain the EoS again, since they are dependent on the constrained EoS results.


\section{Conclusion and Discussion}
\label{sec:discussion}
The measurements of masses of the INSs are challenging tasks. Fortunately, for the objects with a known gravitational redshift, it is possible to reliably estimate the mass, as demonstrated firstly in \citet{Tang2020}. However, in such a work, the possibility of PT was not considered, and the constraints on EoSs mainly come from the nuclear constraints and GW data of GW170817. In the past two years, mass-radius measurements of PSR J0030+0451 and PSR J0740-6620 had been successfully measured, and the radii seem to be larger than those suggested by GW170817. Moreover, the first-order strong PT has been properly incorporated in the parameterizing approach. These progresses motivate us to reestimate the bulk properties (including the masses, the radii, the tidal-deformability and moment of inertia) of a group of INSs. 
 
In comparison to our initial estimates presented in \citet{Tang2020}, the currently inferred gravitational masses of INSs are larger with smaller uncertainties. This is anticipated, because in the EOS constraints, we have added the mass-radius data of two NSs observed by NICER, and both NSs prefer a radius higher than that favored by GW170817 (for a given $z_{\rm g}$, the larger the radius, the higher the gravitational mass). We also find that there is little difference in all the results between PT model and NPT model. This may reflect the fact that most part of these compact objects are not dense enough for the presence of a first-order strong PT. 

NICER is continuing to collect the data to reliably measure the mass-radius of a few more NSs. The LIGO/Virgo/KAGRA network is expected to run in the end of 2022, and many more NS mergers will be detected in the near future. With these data, the constraints on the EoS of NS material will be tightened, and the presence of a first-order strong PT or not will be further probed. These progresses will in turn yield more reliable measurements of the bulk properties of the INSs with known gravitational redshifts.

\acknowledgments
We thank Prof. Yi-Zhong Fan for stimulating discussion. This work was supported in part by NSFC under grant of No. 11773078, 12073080, 11933010, the China Manned Space Project (NO.CMS-CSST-2021-A13), Major Science and Technology Project of Qinghai Province (2019-ZJ-A10), Key Research Program of Frontier Sciences (No. QYZDJ-SSW-SYS024). J.L.J. acknowledges support by the Alexander von Humboldt Foundation.

\software{Bilby \citep[version 1.1.2;][\url{https://git.ligo.org/lscsoft/bilby/}]{2019ApJS..241...27A}, PyMultiNest \citep[version 2.6, ascl:1606.005, \url{https://github.com/JohannesBuchner/PyMultiNest}]{2016ascl.soft06005B}}

\bibliographystyle{aasjournal}
\bibliography{refer.bib}

\end{document}